\documentclass{elsart5p}
\usepackage{graphicx}
\usepackage{amssymb}
\usepackage{amsmath}

\begin{document}

\begin{frontmatter}
\title{Structural distortion in Ludwigites}

\author[aff1]{E. Vallejo\corauthref{cor1}}
\corauth[cor1]{Corresponding author.}
\ead{emmanuel.vallejo@uadec.edu.mx}
\author[aff1]{G. Calderon}
\address[aff1]{Facultad de Ingenier\'{i}a Mec\'{a}nica y El\'{e}ctrica, Universidad Aut\'{o}noma de Coahuila,
C.P. 27276, Torre\'{o}n, Coahuila, M\'{e}xico.}
%\date{\today }

\begin{abstract}
Structural distortion in a three-leg-ladder is studied in connection with Ludwigites, in particular the Fe and Co homometallic ones.
Static impurities in $t_{2g}$-orbitals as infinite repulsion potentials randomly located 
in the three-leg-ladder and a Su-Schrieffer-Heeger like tight-binding Hamiltonian are proposed and discussed. 
It is found that such potentials block itinerant electrons and diminish a structural staggered order parameter, related with structural distortion,
as $3^{-M}$ being M the number of impurities. This diminution is 
in detriment of Peierls like distortion that occurs in these ladders as in the case of Fe-Ludwigite. 
On the other hand, this diminution could explain the lack of structural distortion as in the case of Co-Ludwigite. 
\end{abstract}

\end{frontmatter}

\section{Introduction}\label{}

Interesting physical properties like structural, transport and magnetic ones have been studied in oxyborates compounds known as Warwickites 
\cite{Fernandes1994,Guimaraes1997,Continentino2001} and Ludwigites \cite{Fernandes1998,Guimaraes1999,Fernandes2000,Mir2001,Freitas2009,Bartolome2011,Kazak2011}.
Wigner glass, the existence of a weak ferromagnetism or even charge ordering have been observed in Ludwigite materials 
as consequence of strong electronic correlations and low-dimensional effects. Warwickites are characterized by one-dimensional structures 
called ribbons where the transition metals are randomly located \cite{Fernandes1994,Guimaraes1997}. Ludwigites on the other hand, have a crystalline structure that consists of an 
assembling of subunits. These subunits are in the form of zigzag walls with four nonequivalent octahedral
sites occupied by divalent o trivalent metallic ions \cite{Guimaraes1999}. Warwickites (q=1) and Ludwigites (q=2) present the following chemical formula 
\textit{$M_{q}O_{q}M'BO_{3}$}, where $M(M')$ are divalent (trivalent) $3d$ transition-metal ions.

 M\"{o}ssbauer experiments suggest that Ludwigites, i.e., the homometallic one
$(M=M'=Fe; Fe_{3}O_{2}BO_{3})$ can be viewed as formed of two magnetic systems decoupled to first approximation \cite{Guimaraes1999}. The first one 
are three-leg-ladders (3LL) consisting of triads of $Fe^{3+}$-ions with one itinerant electron per rung. 
The second one are basically 3LL formed by divalent ions. These ions neither participate in charge dynamics nor charge ordering. An antiferromagnetic (AF) 
transition at 112K has been observed in 3LL within the Fe-Ludwigite that basically involves $Fe^{3+}$ and $Fe^{2+}$ 
within the first magnetic system \cite{Larrea2004}. The order is ferromagnetic (F) in the rungs and AF between the rungs along the ladder \cite{Bordet2009}.
The complementary Fe-ions are paramagnetic down to 74K. Below
70K all the sample becomes magnetically ordered. Fe-ions in the second magnetic system order F along the ladder and AF along the rungs \cite{Bordet2009}.
Mostly $Fe^{3+}$-ions order in a weak ferromagnetism where canting of the magnetic 
hyperfine field for each Fe-ion is related to this order \cite{Guimaraes1999,Fernandes2000,Larrea2004}. A total AF state is found
below $\sim40K$ for this Fe-Ludwigite.

Charge ordering at 220K and a structural phase transition at 283K along 3LL 
(first magnetic system) have also been observed by using specific heat measurements \cite{Fernandes2000} 
and X-ray diffraction \cite{Mir2001}, respectively. The structural transition is accompanied by a change in the activation energy of the electrical
resistivity \cite{Mir2001}.  Below 40K two-dimensional AF magnons were proposed to explain the 
low temperature $T^{2}$ behavior and disorder ($Fe^{3+}$-$Fe^{2+}$) was proposed 
to explain the large linear term of the specific heat in $Fe_{3}O_{2}BO_{3}$ above 112K \cite{Fernandes2000}.
Transport measurements in Fe-Ludwigite show activated behavior with two characteristic energies above and below charge ordering at 220K \cite{Guimaraes1999}.
Theoretical investigation of spin exchange interactions and electronic structure were done using a spin-dimer analysis \cite{Whangbo2002} 
and the extended H\"{u}ckel method \cite{Matos2005}, respectively. On the other hand, an excitonic instability using a tight-binding model was studied
in a 3LL \cite{Latge2002}. Nevertheless the important electronic, structural and magnetic correlation was only taken into account till
Ref. \cite{vallejo2006}, where the low temperature $T\lesssim112K$ magnetic phase proposed by neutron powder diffraction study was obtained \cite{Bordet2009}.

On the other hand, at room temperature, the presence of triads of $Co^{3+}$ plus one itinerant electron in 3LL, the crystalline structure, space group, 
bond lengths and lattice parameters of the only other known homometallic oxyborate Co-based Ludwigite ($Co_{3}O_{2}BO_{3}$) 
are very similar to the previous one \cite{Freitas2008}. These similarities suggest similar behaviors, nevertheless their physical properties 
are very distinct \cite{Freitas2008}. Structural transition is not found in the Co-Ludwigite and 
only one AF transition at 42K is found for the former material \cite{Freitas2008}. 
Besides spin-orbit coupling was proposed as a detriment to this Peierls transition inhibiting the structural transition for the Co-Ludwigite \cite{Freitas2008}. 
Furthermore, it was proposed that orbital moments of Co-ions are almost quenched \cite{Bartolome2011}.
Additionally elastic phonon excitations and the role of itinerant
electrons in 3LL could be responsible for the $T^{3}$ and $T$ behavior respectively of the specific heat in Co-Ludwigite. 
On the other hand, at high temperature $T>>42K$ a very complicated thermally activated behavior of the conductivity was found \cite{Ivanova2007}. 
 
In this work, structural distortion in 3LL is studied in connection with homometallic Ludwigites. A disorder-based
mechanism is proposed and analyzed. Impurities as inifinite repulsion potentials are randomly located in these ladders. 
These impurities block itinerant electrons and diminish a structural staggered order parameter, related with structural distortion, as $3^{-M}$ 
where M is the number of impurities. 
This diminution is in detriment of Peierls like distortion observed in these ladders as in the case of Fe-Ludwigite in relationship 
with Co one. The lack of structural distortion in Co-Ludwigite is until now not explained.    

The paper is organized as follows.
In Sec. 2, the mechanism and the tight-binding Hamiltonian is proposed. 
Results of the model are discussed in Sec. 3. Finally conclusions are presented in Sec. 4. 

\section{The mechanism and the tight-binding Hamiltonian}

In this section, a model is introduced to study structural distortion in 3LL 
in order to explain different behaviors 
observed in homometallic Ludwigite systems, i. e., between $Fe_{3}O_{2}BO_{3}$ and $Co_{3}O_{2}BO_{3}$.

At temperature of 283K experimental X-ray diffraction studies show structural transition in 3LL for $Fe_{3}O_{2}BO_{3}$ \cite{Fernandes2000,Mir2001}. 
On the other hand, structural distortion is not observed in $Co_{3}O_{2}BO_{3}$ \cite{Freitas2008}.

Additionally, experimental M\"{o}ssbauer spectroscopy studies determine a structure 3LL consisting of triads of high-spin $Fe^{3+}$-ions
($S=\frac{5}{2}$; $e_{g}^{2}t_{2g}^{3}$) with one itinerant electron per rung for $Fe_{3}O_{2}BO_{3}$. In the case of 
 Co-Ludwigite, 3LL are composed by high-spin $Co^{3+}$-ions ($S=2$; $e_{g}^{2}t_{2g}^{4}$) 
also with one itinerant electron per rung \cite{Freitas2008}.

Furthermore, the transition metal d-orbitals like Fe or Co placed in a crystalline
environment undergo a break of degeneracy due to the crystal-field
resulting in three $t_{2g}$ low-energy orbitals and two $e_{g}$ high-energy ones.

\begin{figure}[h]
\centering  \includegraphics[scale=0.3]{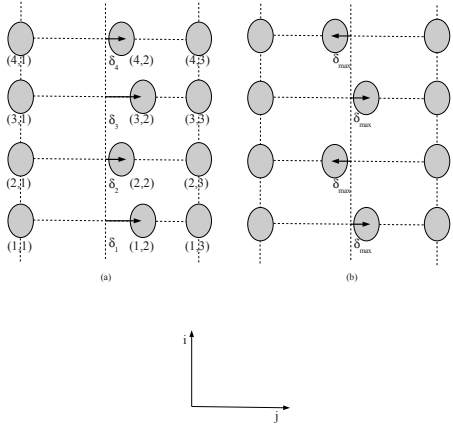}
\caption{(a) Three-leg-ladder proposed in the model. The structural parameters $\delta_{i}$ and $(i,j)$ sites are shown. 
(b) Three-leg-ladder Fe like Ludwigite.}
\label{Fig1}
\end{figure}

To study structural distortion, it is proposed that itinerant electrons move in a linear combination of $t_{2g}$ orbitals in 3LL, see Fig. \ref{Fig1}.
Besides, the electron-lattice coupling is introduced using the Su-Schrieffer-Heeger (SSH) model \cite{Su1979,Su1980} in similar
manner as was done in Ref. \cite{Suarez2009}.

An additional mechanism disorder-based is proposed to analyze the structural distortion in Co-Ludwigite. 
This mechanism is based in impurities randomly placed in sites coming from $t_{2g}$ localized down-electrons of Co-ions. 
These impurities block itinerant electrons and are identified as infinite repulsion potentials.
There is always a probability to have static down-electrons in $t_{2g}$-orbitals very near
to conduction ones and this effect can produce an interaction with itinerant electrons.

To study structural distortion in 3LL a tight-binding Hamiltonian is proposed in Ref. \cite{Latge2002}. 
It is proposed here a more general one which is introduced in the following:
\begin{eqnarray}
\nonumber H &=& -t_{i}\sum_{i,j=1,2,3}(c_{i,j}^{+}c_{i+1,j}+h.c.)-t_{j}\sum_{i}(1+\delta_{i})(c_{i,3}^{+}c_{i,2}+h.c.)\\
\label{eq1}
          &&-t_{j}\sum_{i}(1-\delta_{i})(c_{i,2}^{+}c_{i,1}+h.c.)+B\sum_{i}\delta{i}^{2},
\end{eqnarray}
where $c_{i,j}^{+}(c_{i,j})$ are the spinless fermions creation
(annihilation) operators of the conduction electrons at site $(i,j)$. Index $i$ runs between the rungs along $i$ axis of the ladder
and $j$ corresponds to 1,2 or 3 leg-ladder, see Fig. \ref{Fig1}-(a). 
The variables $t_{i}$ and $t_{j}$,
are hopping parameters between and in the rungs respectively. 
Because of bond lenghts the ratio $t_{j}/t_{i}=1.2$, considered in Ref. \cite{Latge2002} is used. 
To simplify calculations $t_{i}=1$ is set.
The variable B is an elastic energy. Its value is chosen equal to zero which determines 
$\mid\delta_{i,max}\mid\ll1$, specifically it is chosen $\mid\delta_{i,max}\mid=0.2$.\\ 

The parameters $\delta_{i}$ are related with structural distortion, see Ref. \cite{Suarez2009}.  
The lattice distortion is considered in $i$-rungs only 
like experimental distortion results of Fe-Ludwigite, see Fig. \ref{Fig1}-(b).

The Hamiltonian proposed for this model and the disorder-based mechanism are used to analyze structural distortion
in 3LL in connection with homometallic Ludwigites.
\newline

\section{Results and discussion of the model}
In the following, the necessary steps and procedures to calculate observables are explained.

To analyze structural distortion, the energy $\langle U \rangle$, charge density $\langle n_{i,j} \rangle$ and a structural staggered order 
parameter proposed as $d=\frac{3}{N}\mid\sum_{i}(-1)^{i} \langle \delta_{i} \rangle \mid$ are considered.
In the structural staggered order parameter the variable $N$ corresponds to total number of sites in the system 
with a value given by $N=3(i)$, being i the number of rungs in 3LL. 

To obtain observables of the model a set of initial parameters $\delta_{i}$ are randomly chosen. Diagonalization
of this initial Hamiltonian is exactly calculated by standard algorithms. Energy is then obtained. Additionally,
an implementation of a kind of Monte Carlo method is proposed to determine finally the minimum energy. 
After first Monte Carlo simulation a new optimized $\delta_{i}$-configuration is tested. This configuration break the
$\delta_{i}\rightarrow-\delta_{i}$ symmetry. Minimum energy, wave functions and 
$\delta_{i}$-parameters are then obtained. The structural staggered order parameter and charge density are calculated with 
these $\delta_{i}$-parameters and wave functions respectively. 

To simulate the $Co_{3}O_{2}BO_{3}$, impurities are introduced and taken as infinite 
repulsion potentials. They are randomly located in 3LL. An impurity-site configuration and an 
initial $\delta_i$-configuration are randomly proposed to set the initial Hamiltonian.
Minimum energy, wave functions and $\delta_{i}$-parameters are calculated as before.
 
Mean values of these observables, i. e., $\langle U \rangle = \sum{P_{i}U_{i}}$ can be calculated over all possible
impurity-site configurations, it means over disorder. For simplicity, each impurity-site configuration has the same probability.
On the other hand, closed (CBC) and periodic (PBC) boundary conditions are used.
Besides an electron-only approximation is used.

\subsection{zero impurities $Fe_{3}O_{2}BO_{3}$ Ludwigite case}

For the case without impurities initial results of Monte Carlo simulations are presented. Zero impurities mean 
$\langle \delta_{i} \rangle=\delta_{i}$ or not disorder. An optimized zigzag-configuration with values $\mid\delta_{i}\mid\rightarrow \delta_{max}$
is always obtained for every Monte Carlo calculation. This configuration can be observed in Fig. \ref{Fig1}-b.

\begin{figure}[h]
\centering  \includegraphics[scale=0.8]{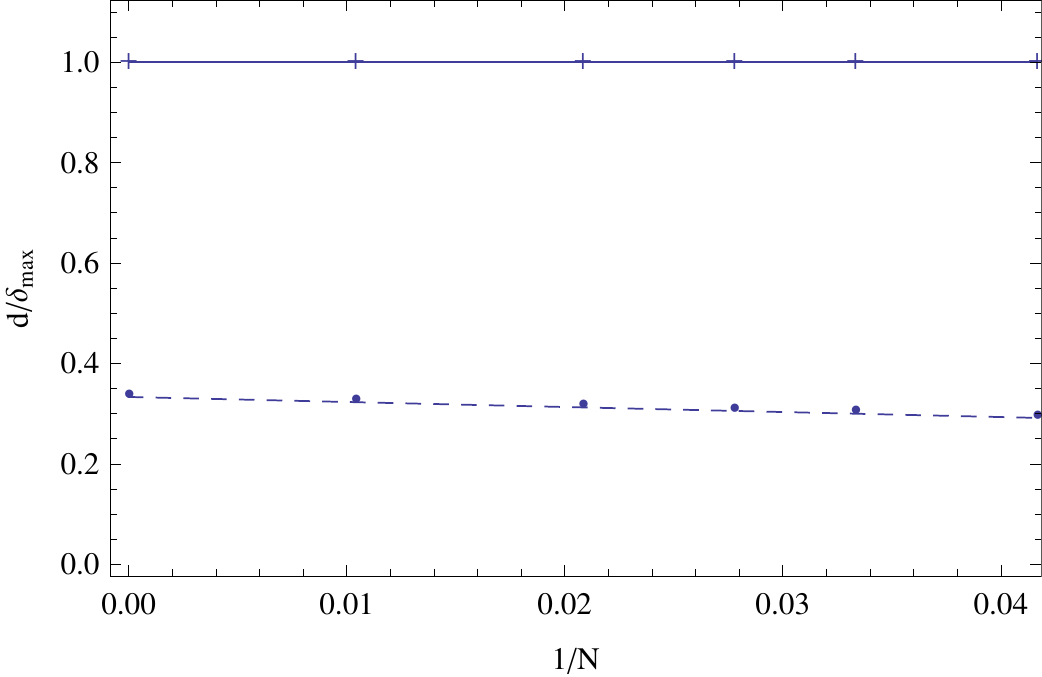}
\caption{Structural staggered order parameter $d/\delta_{max}$ vis 1/N sites. CBC are presented in this case. Full and dashed lines 
represent zero and one impurity respectively. Lines are only guides for the eyes. The thermodynamic limit ($N\longrightarrow\infty$) is also included.}
\label{Fig2}
\end{figure}

In this case structural staggered ordered parameter gives $d/\delta_{max}=1$, see continuous line in Fig. \ref{Fig2}. 
This result is in agreement with experimental results reported for Fe-Ludwigite \cite{Mir2001}.

The 3LL charge density is shown in Fig. \ref{Fig3}, see continuous line. 
In this case structural distortion and charge ordering are observed and they are also correlated.
 
The optimized structure proposed in Monte Carlo simulations can be observed in Fig. \ref{Fig1}-(b). 
Charge density in (1,3)-site is larger than (1,1)-site because structural distortion.     

This zigzag-distortion is also obtained using
PBC like in Ref. \cite{Latge2002}. For the former boundary condition 
a gap is open lowering energy for n=1/3 itinerant band filling. 

For CBC normalized energy $\langle U \rangle/N$
as a function of 1/N sites is shown in Fig. \ref{Fig4}. This energy is almost N-site independent. 
\subsection{finite impurities $Co_{3}O_{2}BO_{3}$ Ludwigite case}
\begin{figure}[h]
\centering  \includegraphics[scale=0.8]{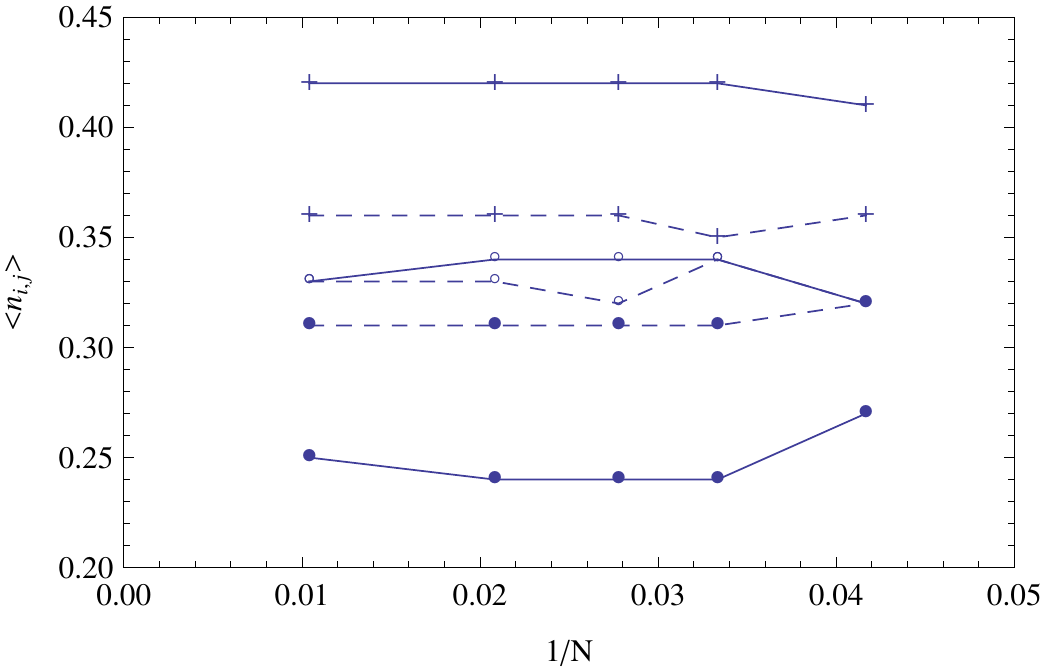}
\caption{The 3LL charge density $\langle n_{i,j} \rangle$ vis 1/N sites. Full and dashed lines correspond to zero and one impurity respectively. These lines are guides
for the eyes. Plus, open and filled circles symbols are $\langle n_{1,3} \rangle$, $\langle n_{1,2} \rangle$ and $\langle n_{1,1} \rangle$ charge density following Fig. \ref{Fig1}.}
\label{Fig3}
\end{figure}

\begin{figure}[h]
\centering  \includegraphics[scale=0.8]{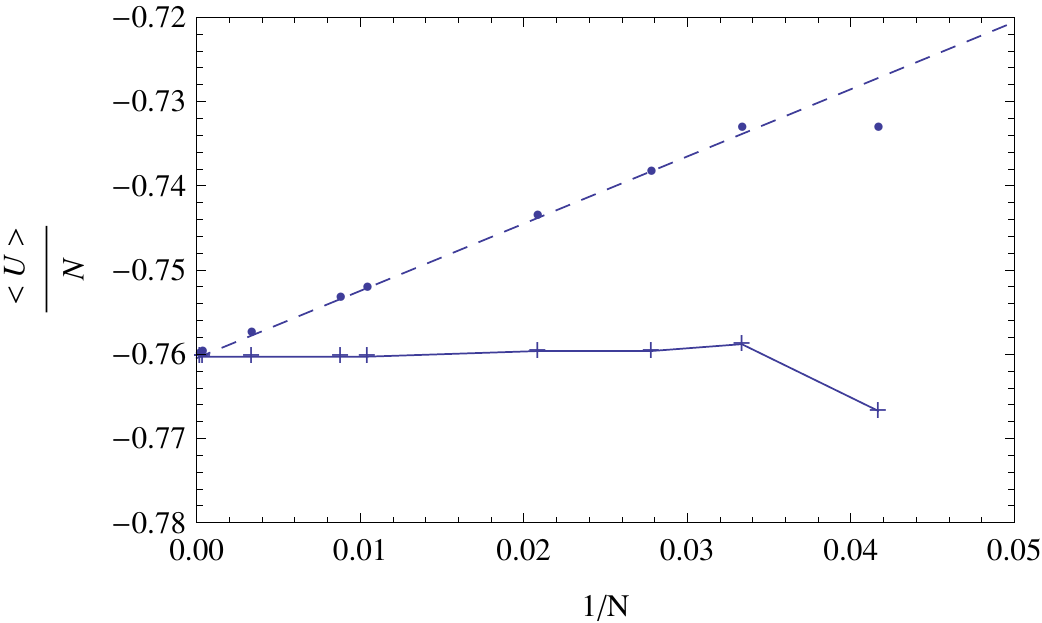}
\caption{The 3LL mean energy $\langle U \rangle/N$ vis $1/N$ sites. Full and dashed lines represent zero and one impurity respectively. Dashed line is a least
squares fitting. Lines are guides for the eyes.}
\label{Fig4}
\end{figure}

The structural staggered order parameter is considerably diminished in presence of one impurity as can be seen in Fig. \ref{Fig2}.
Because of symmetry reasons, contributions to the average of the structural parameter are cancelled when the impurity is located in sites where (i,1)
$\delta_{i}=\delta_{max}$ and (i,3) $\delta_{i}=-\delta_{max}$, respectively.
In this case, the contribution to $\langle \delta_{i} \rangle$ is given when the impurity is located in the middle site (i,2), being $\delta_{i}=0$. 
For all these contributions, the same zigzag structural order like Fe-Ludwigite (without impurities) is found from Monte Carlo calculations, see Fig. 1-(b). 
Again an initial zigzag condition and $\mid\delta_{i}\mid=\delta_{max}$ are proposed, as an initial optimized configuration, and there are always obtained for every Monte Carlo simulation.
When the impurity is located in the middle site a gap is open like Fe-Ludwigite previously studied \cite{Latge2002} as can be seen by mean of PBC results.  
It is found that $d/\delta_{max}=1/3-1/N$ (straight dashed line in Fig. \ref{Fig2}). The thermodynamic limit ($N\rightarrow\infty$) corresponds to
$d/\delta_{max}=1/3$. One impurity diminishes a third part of the structural staggered order parameter $d/\delta_{max}$.\\
\newline
This zigzag structural ordering leads also to a zigzag charge ordering as can be seen in Fig. \ref{Fig3},  
zero to compare and one impurity case. The first rung is only shown in Fig. \ref{Fig3} because zigzag symmetry reasons. It means $\langle n_{i,1} \rangle \cong \langle n_{i+1,3}\rangle$,
$\langle n_{i,3} \rangle \cong \langle n_{i+1,1}\rangle$ and $\langle n_{i,2} \rangle \cong \langle n_{i+1,2} \rangle$. 
The electronic charge moves $(i,1)\Leftrightarrow(i,3)$. It is clear that if $\delta_{max}\rightarrow0$ then 
$\langle n_{i,1(3)}\rangle \leftrightarrow \langle n_{i,3(1)} \rangle$.
\bigskip
%\begin{center}
%\end{center}
\newline
The mean energy $\langle U \rangle/N$ vis $1/N$ sites for 3LL can be observed in Fig. \ref{Fig4}. 
Dashed line corresponds to a least squares fitting given by 
\begin{equation}
\langle U\rangle /N = (-0.76\pm 0.14\times 10^{-3})+(0.79\pm 0.79\times 10^{-2})/N.
\end{equation}
The thermodynamic limit $N\rightarrow\infty$ implies
$\langle U \rangle/N=-0.76\pm0.14\times10^{-3}$. This value can be compared with PBC for zero impurities $\langle U \rangle/N=-0.76$ (value observed in the thermodynamic limit, see Fig. \ref{Fig4}).
PBC and CBC give the same energy in the thermodynamic limit. Both energies, for zero and one impurity, are very closed 
because a single impurity is difficult to see in a thermodynamic limit. Additionally if the system is not too large the impurity clearly modifies mean energy as observed in Fig. \ref{Fig4}. 
\bigskip
\newline
In general, because of mirror symmetry reasons (in the middle site (i,2)), the most important contribution to the structural staggered order
parameter is when impurities are located in the middle site (i,2) of the 3LL. The structural staggered order parameter is obtained as 
\begin{equation}
\label{eq2}
 \frac{d}{\delta_{max}}=\frac{\frac{(\frac{N}{3}-1)!}{M!(\frac{N}{3}-M-1)!}}{\frac{N!}{M!(N-M)!}}+\epsilon(\frac{1}{N}),  
\end{equation}
where M is the number of impurities and $\epsilon$ is the error for taking only middle site for impurities. 
For zero and one impurity $\epsilon$ equals exactly to zero and the previously results are recovered, i. e. $d/\delta_{max}=1$ and  
$1/3-1/N$, respectively. For $M>1$, the result $\epsilon\rightarrow0$ is obtained such as the thermodynamic limit is reached. 
For example, if M=2 (two impurities) there are two different contributions to $\langle \delta \rangle$ because of symmetry reasons. 
The first one is when impurities are located in sites (i,1) and (i,3).
The second one for impurities placed in (i,2) and (j,2) being $i\neq j$ (middle site). If $N>>6$ the first contribution is proportional to N/3 and the second one to
$N^{2}/(2(3)^{2})$. There are $N^{2}/2$ total contributions. If we are only interested in the thermodynamic limit 
the former gives $d/\delta_{max}=1/3^{2}$. Almost one magnitude order decreases from zero impurities results. \\
\newline
In the thermodynamic limit, Eq. (\ref{eq2}) can be written as $d/\delta_{max}=1/3^{M}$. The former equation can be
rewritten as $d/\delta_{max}=1/3^{x_{i}N}$, being $x_{i}$ concentration impurities.
For a finite concentration impurities, then it is found the limit $d/\delta_{max}\rightarrow0$.  \\
\newline
Because of impurities, the structural staggered order parameter $d/\delta_{max}$ 
diminishes as $1/3^{M}$ and could be near to the experimental error of X-ray diffraction studies or temperature ion-position fluctuations.  
Experimental error of X-ray diffraction studies correspond to $\sim 10^{-4}(10^{-5})$ Ref. \cite{Mir2001,Freitas2008}. In this case the number
of impurities can be fixed. For the impurity-condition given by $M\geq12$, the structural staggered order parameter has a value given by 
$d/\delta_{max}\leq\sim10^{-6}$.
For this impurity-condition structural distortion can not be identified.  
This result could be related with
experimental results of Co-based Ludwigite \cite{Freitas2008}. It is necessary to consider that this Ludwigite does not show structural transition. 
Static impurities or infinite repulsion potentials as proposed here, could be responsible for this lack of structural transition in this material.
 On the other hand, the Fe-based Ludwigite and the zigzag structural distortion, experimentally observed, are recovered using 
the number of impurities M=0.\\
\newline
Next, it is discussed this mechanism in connection with other oxyborate systems. 
For example, at temperatures of 
293K and 150K, X-ray diffraction studies of heterometallic oxyborates, i. e., 
$Co_{2}FeO_{2}BO_{3}$ and $Ni_{2}FeO_{2}BO_{3}$ do not show structural transition, see Ref. \cite{Freitas2009}. 
Additionally, structural and magnetic properties of the oxyborate $Co_{5}Ti(O_{2}BO_{3})_{2}$ were investigated \cite{Freitas2010}. 
For this compound, X-ray diffraction measurements at 293K and 150K do not show either structural transition \cite{Freitas2010}. 
Static impurities come from Co or Ni in $t_{2g}$-orbitals could block the
structural transition in those materials as suggested in this work.\\

\section{Conclusions}
A mechanism disorder-based and a tight-binding Hamiltonian are used and discussed to study structural distortion in 
three-leg-ladders in connection with a kind of oxyborates known as Ludwigites. In particular the Fe and Co ones. 

It is found that static impurities as infinite repulsion potentials come from Co or other transition metals like Ni 
can block the itinerant electrons and diminish a structural staggered order parameter, related with structural distortion, as $1/3^{M}$ being M the number of impurities. 
This diminution could be comparable with experimental error of X-ray diffraction studies or temperature ion-position fluctuations and could explain this lack
of structural distortion in for example the Co-Ludwigite case. 
Experimental error of X-ray diffraction studies correspond to $\sim 10^{-4}(10^{-5})$. In this case an
impurity-condition given by $M\geq12$ can be fixed, where structural distortion can not be observed. 
This lack of structural distortion is until now not explained for Co-Ludwigite.

A zigzag structural distortion is always found in all simulations.
The particular structural distortion obtained for $M=0$ (zero impurities) 
is directly related with experimental X-ray diffraction studies for Fe-Ludwigite.  

This mechanism is discussed in connection with other oxyborate systems, i.e. $Co_{2}FeO_{2}BO_{3}$, $Ni_{2}FeO_{2}BO_{3}$ and 
$Co_{5}Ti(O_{2}BO_{3})_{2}$, where structural distortion is not observed either. In those cases impurities
come from Co or Ni in $t_{2g}$-orbitals could block the
structural distortion as suggested in this work.

\begin{center}
\textbf{ACKNOWLEDGMENT}
\end{center}
This work was partially supported by Programa de Mejoramiento de Profesorado PROMEP UACOAH-PTC-238.

\end{document}